\documentclass[a4paper,12pt]{article}

\usepackage{amsmath}
\usepackage{amssymb}
\usepackage{graphicx}
\usepackage{hyperref}

\numberwithin{equation}{section}

\def\tr{\mathrm{tr}}
\def\SO{\mathrm{SO}}
\def\SU{\mathrm{SU}}

\def\U{\mathrm{U}}
\def\cN{\mathcal{N}}
\def\bC{\mathbb{C}}
\def\bR{\mathbb{R}}
\def\bZ{\mathbb{Z}}

\begin{document}

\begin{titlepage}
\begin{flushright}
\normalsize
{\tt IPMU-11-0014}

\medskip

February, 2011
\end{flushright}
\vfil

\bigskip

\begin{center}
\Large On W-algebras 
and the symmetries \\
of  defects of 6d $\cN=(2,0)$ theory 
\end{center}

\vfil
\medskip

\begin{center}
\def\thefootnote{\fnsymbol{footnote}}

 Yuji Tachikawa\footnote[1]{on leave from the Institute for Advanced Study} 

\bigskip

Institute for the Physics and Mathematics of the University,

\medskip

University of Tokyo, Kashiwa, Chiba 277-8583, Japan

\end{center}

\vfil
\bigskip

\begin{center}
{\bfseries abstract}
\end{center}

\bigskip

We consider 6d $\cN=(2,0)$ theory on $N$ M5-branes,
together with a 4d defect labeled by a Young diagram $Y$ specifying its global symmetry $G_Y$.
A recent conjecture states that a compactification of this system leads to a 2d theory with W-algebra symmetry depending on  $Y$.
We provide a check of the conjecture by
reproducing the level of the current subalgebra $\hat G_Y$ of this W-algebra 
from the property of the 4d defect.

\vfill

\end{titlepage}

\section{Introduction}
The low-energy limit of $N$ coincident M5-branes (minus the center-of-mass mode)  is the mysterious 6d $\cN=(2,0)$ theory of type $A_{N-1}$.
A nice class of codimension-two defects of this 6d theory was described in \cite{Gaiotto:2009we}.
Within this class, the type of the defect is characterized by a Young diagram $Y$ with $N$ boxes.
The subject of the note is to study the global symmetry on this defect.

Put the $N$ M5-branes along the directions 012345 in a flat 11d spacetime,
and let a defect of type $Y$ fill the directions 0123:\footnote{The brane configuration here shows the defect as created by an intersecting M5-brane. The defect can also come from Taub-NUT spaces.  and also from a combination of M5-branes and NUT singularities. The structure of the breaking of $\SO(5)_R$ symmetry is always the same, and the brane configuration such as \eqref{construction} should be interpreted as a shorthand showing the R-symmetry structure by a given defect. }
\begin{equation}
\begin{array}{l|cc|cc|cc|cc|ccccc}
&0&1&2&3&4&5&6&7&8&9&10 \\
\hline
\hbox{$N$ M5s} & - & - & -& -& - & - &&&&  \\
\hbox{defect} & - & - & -& -&  &  & - &- &&\\
\end{array} \label{construction}
\end{equation}
Here, the sign $-$ in the column labeled by $i$ means that the particular brane extends along the $i$-th direction $x_i$. 
The rotation of the directions 45 and that of the directions 67 act on the defect as the global symmetry. We call the former $\SO(2)_T$ (because the directions are transverse to the defect inside the 6d theory)
and the latter $\SO(2)_R$ (because it becomes the R-symmetry of the combined 4d $\cN=2$ theory.)

In addition, a subgroup $G_Y\subset \SU(N)$ is known to arise as a global symmetry localized on the four-dimensional defect. 
Then the 't Hooft anomaly of the defect includes the following: \begin{equation}
\SO(2)_{T}  G_Y  G_Y \quad \text{and}\quad
\SO(2)_R G_Y G_Y.
\end{equation} 
They manifest themselves in different guises in  distinct situations.
\begin{itemize}
\item The first is in the 4d $\cN=2$ theory obtained by compactifying the directions 45 on a Riemann surface with a suitable twist to preserve supersymmetry \cite{Gaiotto:2009we}. 
Then  the  $\SO(2)_R G_YG_Y$ anomaly gives the two-point function of currents of $G_Y$.
\item The second is in the gravity dual of this system, studied in \cite{Gaiotto:2009gz}. On the gravity side, there are gauge fields associated to $\SO(2)_{T}$,
$\SO(2)_{R}$ and $G_Y$,
and the anomalies are encoded by  Chern-Simons couplings among them.
\item The third is the 2d theory obtained by compactifying the directions 2345 on the omega-deformed flat space $\bR^4_{\epsilon_1,\epsilon_2}$,  again with an appropriate twist to preserve supersymmetry.
When the Young diagram consists of a row of $N$ boxes, this 2d theory is known to have $\hat{\SU}(N)$ current algebra symmetry \cite{Alday:2010vg,Kozcaz:2010yp}. 
For a general Young diagram $Y$, it was conjectured \cite{Braverman:2010ef,Wyllard:2010rp,Wyllard:2010vi} to have W-algebra symmetry obtained by the quantum Drinfeld-Sokolov reduction of $\hat\SU(N)$ algebra with respect to an $\SU(2)$ embedding $\rho_Y:\SU(2)\to \SU(N)$ given by the Young diagram $Y$.
This general W-algebra has  $\hat{G}_Y$ current algebra as a subalgebra.
Then, the level of this $\hat{G}_Y$ current algebra should be given by a linear combination of the anomalies of the original 4d defect.
\end{itemize}

The aim of our note is, then, to calculate these anomalies  in each description, and show that they are related as dictated by the choice of the compactifications used. We will see that, indeed, the current algebra levels are indeed reproduced by this procedure, thus providing a small check of the recent conjecture that the general W-algebra lives on a compactification of the general defect.

The rest of the note is organized as follows:
In section~\ref{5}, we quickly recall how the Young diagrams characterize the codimension-two defects, mainly to establish our notation for the Young diagrams.
In section~\ref{2}, we  review the construction of the W-algebra of type $Y$,
and determine the levels of the affine algebras contained in it. 
In section~\ref{4}, we recall the construction of the 4d $\cN=2$ quiver gauge theories for the defect of type $Y$ and the determination of the current two-point functions of $G_Y$.
In section~\ref{11}, we review the 11d supergravity for the defect of type $Y$ and read off the anomaly coefficients of $G_Y$.
In section~\ref{conclusions}, we show that the levels determined in section~\ref{2}
and the current two-point functions in section~\ref{4} are the consequences of the anomaly polynomial of the defects determined in section~\ref{11}.
We then conclude the note with a short discussion.
In an appendix, we describe the supersymmetry of the configurations we use in the paper and those used in the previous papers in more detail.

As the reader would surely notice, all of the essential calculations had  already been done in the papers cited in the corresponding sections. 
Our point is that all of them can be uniformly understood from the properties of the codimension-two defects of the 6d $\cN=(2,0)$ theory.
Superstring/M theory has shown a capacity to absorb every theoretical construct; now is the turn of W-algebras to be so incorporated.

\section{Defects and Young diagrams}\label{5}

A codimension-two defect of the 6d theory can give rise to many types of defects in the lower-dimensional theory obtained by compactifications.
For example, when the system is compactified along $T^2$ parallel to the defect, it is expected to become a codimension-two defect of 4d $\cN=4$ Yang-Mills theory studied in \cite{Gukov:2006jk}. For more on this point, see e.g.~Sec.~6.4 of \cite{Witten:2011zz}.

The relation of the Young diagram to the global symmetry  becomes clearer when we put the system on a cigar geometry with the defect at the tip, and compactify the $S^1$ around the defect. 
Then we have a codimension-one boundary of 5d maximally-supersymmetric $\SU(N)$ Yang-Mills theory \cite{Gaiotto:2009hg}, which can be analyzed much as in \cite{Gaiotto:2008sa}.
Three adjoint scalars $\phi_{1,2,3}$ (out of five) become singular as we approach the boundary,
breaking $\SO(5)_R$ down to $\SO(2)_R\times \SO(3)_R$.
Their evolution transverse to the boundary follows the Nahm equation.
As such, for  a given triplet $t_1$, $t_2$ and $t_3$ of  $N\times N$ traceless matrices satisfying $\SU(2)$ commutation relations, we can specify the local form of the fields: \begin{equation}
\phi_i  \sim  {t_i}/{y}
\end{equation} where $y$ is the distance to the boundary.

\begin{figure}
\[
\vcenter{\hbox{\includegraphics[width=.2\textwidth]{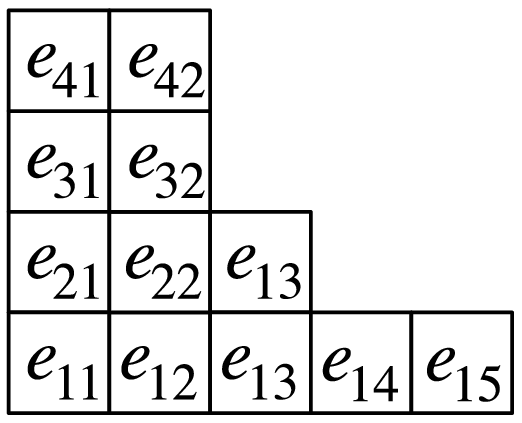}}}
\qquad
\begin{array}{r@{}l}
Y=(n_i)&=(4,4,2,1,1),\\
(m_i)&=(5,3,2,2),\\
(\eta_i^{\ell_i})&=(4^2,2^1,1^2), \\
(\lambda_i)&=(5,8,10,12,12,\ldots).
\end{array}
\]
\caption{A Young diagram and various sequences associated to it. \label{young} }
\end{figure}

Such a triple of $t_i$'s gives an $N$-dimensional representation of $\SU(2)$, \begin{equation}
\rho: \SU(2) \to \SU(N)
\end{equation} and is characterized by the decomposition of the $N$-dimensional representation of $\SU(N)$ into the irreducible representations of $\SU(2)$: \begin{equation}
\underline{N}\to \underline{n_1} \oplus \underline{n_2} \oplus \cdots  
\end{equation} where $\underline{n}$ is the $n$-dimensional irreducible representation of $\SU(2)$, and we order such that $n_i\ge n_j$ when $i>j$. 
Then $(n_i)$  determines a partition of $N$, or equivalently a Young diagram $Y$.
We identify the partition and the diagram, and often write $Y=(n_i)$.
One example is shown in Figure~\ref{young}.
We define integers $\eta_i$, $\ell_i$ such that \begin{equation}
(n_i) = (\underbrace{\eta_1,\ldots,\eta_1}_{\ell_1\ \text{times}}, \underbrace{\eta_2,\ldots,\eta_2}_{\ell_2\ \text{times}}, \ldots).
\end{equation}

The $\SU(N)$ gauge rotation on the boundary is not part of the local gauge symmetry but the global symmetry. It is, however, explicitly broken by the choice of $\rho$. Then, 
what remains as the global symmetry of the defect is the maximal commuting subgroup $G_Y$ to $\rho(\SU(2))$: \begin{equation}
\rho(\SU(2))\times G_Y \subset \SU(N).
\end{equation}
This $G_Y$ has the form \begin{equation}
G_Y = ( G_1\times G_2 \times \cdots )/\U(1) \quad \text{where} \quad G_i=\U(\ell_i).
\end{equation} where the quotient is by the diagonal $\U(1)$ subgroup of all $G_i$.
We denote the dual partition to $(n_i)$ by $(m_i)$, \begin{equation}
m_i = \# \{ j \  | \  n_j \ge i \}.
\end{equation}
We will also need the sequence $(\lambda_i)$, defined by \begin{equation}
\lambda_i = \sum_{j\le i} m_j.
\end{equation} Here the range of $i$ is taken from one to infinity.

\section{W-algebras}\label{2}

Let us begin this section by describing the relation of the construction \eqref{construction}  of the 2d theory in the introduction to the more conventional one emphasizing the instanton integral.\footnote{This better point of view explained here was suggested to the author by J\"org Teschner during the summer workshop at Santa Barbara in 2010.}

Compactify the direction 1 in the construction \eqref{construction} on an $S^1$. 
Then the 6d theory becomes 5d maximally-supersymmetric $\SU(N)$ Yang-Mills theory with a codimension-two defect operator. 
Taking the direction 0 as the time direction and considering only the BPS sector, the system reduces to the supersymmetric quantum mechanics on the moduli space of instantons on $\bR^4$ spanned by directions 2345, with a suitable singularity at $x_4=x_5=0$ dictated by the type of the defect. 
Under the twist by $\epsilon_{1,2}$, the zero-energy states are the equivariant cohomology of the moduli space, and the W-algebra symmetry we describe below will act on it.

For $Y=(N)$, the defect does not add local degrees of freedom.
The theory is effectively without the defect operator, and the moduli space is the standard moduli space of instantons. 
Then the 2d theory has the standard $W_N$-symmetry, as discussed in  \cite{Alday:2009aq,Wyllard:2009hg}. 
Note that we only talked about 5d gauge theory without the compactified $S^1$ of the original system. 
But the analysis of the BPS sector of the 5d theory regenerates the compactified $S^1$, with $L_0$  of the 2d theory (which is more or less the KK momentum along the $S^1$) identified with the instanton number, as is standard in the compactification of the $\cN=(2,0)$ theory.\footnote{This regeneration of $S^1$ from the 5d point of view was recently discussed in \cite{Douglas:2010iu,Lambert:2010iw}.}

For $Y=(1,1,\ldots,1)$, the defect has the full $\SU(N)$ symmetry,
and the 2d theory has the $\hat\SU(N)$ affine symmetry \cite{Alday:2010vg,Kozcaz:2010yp}.
Our focus is on the defects whose type is between these two extremes, $Y=(N)$ and $Y=(1,\ldots,1)$.
In the following, we need a few properties of the general W-algebras. For extensive references, see the review \cite{Bouwknegt:1992wg} and the reprint volume \cite{Bouwknegt:1995ag}. 

General W-algebras are obtained from the quantum Drinfeld-Sokolov reduction applied to an affine Lie algebra $\hat G$ \cite{deBoer:1993iz}. 
The procedure involves a choice of an embedding  $\rho:\SU(2)\to G$,
and gauge-fixes in terms of the BRST cohomology
the part of $\hat G$ which does not commute with the image of $\rho$.
We denote the resulting algebra by $W(\hat G,\rho)$. This contains the affine Lie algebra $\hat G_\rho$ as a subalgebra, where $G_\rho$ is the maximal commuting subgroup to $\rho(\SU(2))$ inside $G$.

Let us specialize to the case $G=\SU(N)$.
As in the previous section, we associate a representation $\rho_Y$ of $\SU(2)$  for a Young diagram $Y$ with $N$ boxes.
$\rho_Y$ is trivial for $Y= (1,1,\ldots,1)$. Then $W(\hat \SU(N),\rho_Y)$ is just the original $\hat \SU(N)$ affine Lie algebra. 
The standard $W_N$ algebra \cite{Zamolodchikov:1985wn,Fateev:1987zh} is obtained by taking  $Y=(N)$ \cite{Bershadsky:1989mf,FigueroaO'Farrill:1990dz,Feigin:1990pn}.  

In general, $W(\hat\SU(N),\rho_Y)$ can be realized as a subalgebra of the affine Lie algebra $\hat G'_Y$ \cite{deBoer:1993iz}: \begin{equation}
\hat G_Y \subset W(\hat\SU(N),\rho_Y) \subset \hat G'_Y\label{Miura}
\end{equation} where the inclusions are as vertex operator algebras.
Here \begin{equation}
G'_Y = [G'_1 \times G'_2 \times\cdots ] / \U(1), \quad G'_i = \U(m_i).
\end{equation} 
The level of $\hat G'_i \subset \hat G'_Y$ is given by $k+(N-m_i)$, where $k$ is the level of the original $\hat\SU(N)$ algebra to which the quantum Drinfeld-Sokolov reduction is applied \cite{deBoer:1993iz}.
For $Y=(1,\ldots,1)$, the group $G'_Y$ is $\U(1)^{N-1}$, and the embedding \eqref{Miura} expresses  the $W_N$-algebra in terms of $N-1$ free bosons; this is called the quantum Miura transformation.

The relation of groups\footnote{Note that this does \emph{not} mean that $\hat G'_Y\subset \hat G$; the levels are different. }\begin{equation}
G_Y \subset G'_Y \subset G=\SU(N)
\end{equation} is described using the Young diagram $Y$.
We associate a basis vector $e_{ij}$ for a box at position $(i,j)$ in the Young diagram $Y$, see Fig.~\ref{young}. We have in total $N$ basis vectors.
Then $G=\SU(N)$ rotates all the basis vectors,
and under the $\SU(2)$ representation $\rho_Y$,
the vectors $e_{1i}$, \ldots, $e_{n_i i}$ form an irreducible representation of dimension $n_i$.
$G'_i\subset G'_Y$ rotates the basis vectors $e_{i1}$, $e_{i2}$, \ldots, $e_{im_i}$.
$G_i \subset G_Y$ then acts on the basis vectors by permuting the columns with height $\eta_i$,
commuting with $\rho_Y$. 
In other words, $G_i=\SU(\ell_i)$ is a diagonal subgroup of all $\SU(\ell_i)\subset G'_j$ 
such that $j\le \eta_i$.

This immediately fixes the level $k_i$ of $\hat G_i\subset W(\hat\SU(N),\rho_Y)$:
\begin{align}
k_i = \sum_{j=1}^{\eta_i} \left[ k+(N-m_i) \right] = - \lambda_{\eta_i} - \eta_i b^2\label{2dlevel}
\end{align} where we introduced the parameter $b$ via  $k=-N-b^2$, as is done conventionally.
Therefore, we expect to find this level upon the compactification of the codimension-2 defect as described in the introduction.

\section{4d $\cN=2$ gauge theory}\label{4}

\begin{figure}
\[
\includegraphics[width=.6\textwidth]{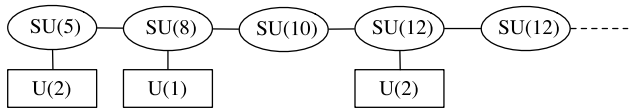}
\]
\caption{Quiver tail corresponding to the Young diagram in Fig.~\ref{young}.  \label{quiver} }
\end{figure}

Given a Young diagram $Y$, we associate, following \cite{Gaiotto:2009we}, a semi-infinite quiver gauge theory of the form \begin{equation}
\SU(\lambda_1)\times \SU(\lambda_2)\times \cdots
\end{equation} where there are  hypermultiplets in the bifudamental representation of $\SU(\lambda_i)\times \SU(\lambda_i)$ for each $i$, and $\ell_i$ copies of the fundamental representation of $\SU(\lambda_i)$. 
The beta function of each of the gauge groups  vanishes. 
This system can be realized in Type IIA and lifted to M-theory, resulting in $N$ M5-branes wrapped on a Riemann surface, with a number of codimension-two defects. 
One is of type $Y$, and carries the flavor symmetry of the fundamentals, which is \begin{equation}
G_Y=[\U(\ell_1)\times \U(\ell_2)\times \cdots ] / \U(1).
\end{equation}
The other defects are of type $(n_i)=(N-1,1)$, each of which carries a $\U(1)$ flavor symmetry, which is identified with the flavor symmetry of a bifundamental. See Fig.~\ref{quiver} for an example.

Then the two-point function of the global symmetry current for $\SU(\ell_i)$ is given by $\lambda_{\eta_i}$. Or equivalently, the anomaly  the 't Hooft anomaly $\U(1)_R\SU(\ell_i)^2$  is characterized by the anomaly polynomial 
\begin{equation}
\sum_i \lambda_{\eta_i} \frac{F_R}{2\pi} \wedge \frac{1}{8\pi^2}\tr F_i \wedge F_i.\label{4danomaly}
\end{equation} 
Here, $F_R$ and $F_i$ are the external gauge fields coupled to the R-symmetry and $\SU(\ell_i)\subset G_i$, respectively,
and we normalized so that the R-charge of the fermion component of a free hypermultiplet is 1/2. 
Each hypermultiplet contains two Weyl fermions, giving the coefficients shown in \eqref{4danomaly}.

\section{Supergravity description}\label{11}
\begin{figure}
\[
\includegraphics[width=.4\textwidth]{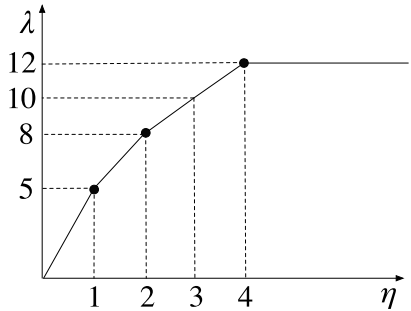}
\]
\caption{Graph of $\lambda(\eta)$ for the Young diagram given in Fig.~\ref{young}. The slope of the function changes discontinuously at the black points shown.\label{graph}}
\end{figure}

The gravity dual solution for a defect of type $Y$ was constructed in \cite{Gaiotto:2009gz}, following the methods of \cite{Lin:2004nb,Lin:2005nh,OColgain:2010ev}. The solution follows from the general ansatz of half-BPS geometries of 11d supergravity, which reduces the full equations of motion to the 3d Toda equation. The rotational symmetry around the defects then reduces the 3d Toda equation to the axisymmetric electrostatic problem for the potential $V$ in the 3d space with the vertical direction $\eta$ and the plane whose radial coordinate is $\rho$.
Introducing $\dot V\equiv \rho\partial_\rho V$ and $V'\equiv \partial_\eta V$, the Laplace equation is \begin{equation}
\ddot V+\rho^2 V''=0
\end{equation} so that there is a line charge at $\rho=0$ given by \begin{equation}
\dot V(\eta)_{\rho=0} = \lambda(\eta).
\end{equation}
For a Young diagram $Y$, the function $\lambda(\eta)$ is given by connecting  by linear segments the points $(i,\lambda_i)$ as defined in section~\ref{5}, see Fig.~\ref{graph} for an illustration.

The supergravity solution close to $\rho=0$ is then given by  \cite{Gaiotto:2009gz} \begin{align}
ds^2 &\sim  \kappa^{2/3} \left(\frac{\dot V\tilde \Delta}{2V''}\right)^{1/3}\Bigl[
4 ds^2_{AdS_5} + \frac{2V''\dot V}{\tilde \Delta}ds^2_{S^2} + 
 \nonumber \\
& \qquad \frac{2V''}{\dot V}(d\rho^2+\rho^2d\chi^2 + d\eta^2) +\frac{4}{\tilde \Delta}(d\beta +\dot V'd\chi)^2\Bigr], \\
\tilde \Delta &\sim 2 \dot V V'' + (\dot V')^2 ,\\
C_3 &\sim \frac{1}{8\pi^2} \left[
(-\dot V+\eta\dot V')d\chi+(\frac{\dot V\dot V'}{\tilde \Delta}-\eta) (d\beta+\dot V'd\chi)
\right] d\Omega_2. \label{C-field}
\end{align} 
Here, $C_3$ is normalized so that it couples to the M2 brane via the phase $\exp(2\pi i\int C_3)$.
The coordinates $\beta$ and $\chi$ have periodicity $2\pi$. At $\rho=0$, the $\chi$ direction shrinks, but the direction $\beta+\dot V'\chi$ remains of finite radius. This quantizes the slope of $\lambda(\eta)$ to be integers. 

This slope changes discontinuously at $\eta=\eta_i$: \begin{equation}
\dot V''(\eta)_{\rho=0} = \lambda(\eta)''=-\sum_i \ell_i \delta(\eta-\eta_i).
\end{equation} 
At $\eta=\eta_i$, both the directions $\chi$ and $\beta$ shrink, 
creating the space of the form $\bC^2/\bZ_{\ell_i}$.
There, the $C_3$ field \eqref{C-field} can be approximated by \begin{equation}
C_3\sim \frac{1}{8\pi^2}(-\lambda_{\eta_i} d\chi - \eta_i d\beta) d\Omega_2.
\end{equation}
When $\ell_i>1$  there is an $\SU(\ell_i)$  gauge symmetry emergent on the locus of the singularity,
and there is the M-theory coupling \begin{equation}
\sim \int C_3\wedge \frac{1}{8\pi^2}\tr F_i\wedge F_i,\label{CS}
\end{equation} where $F_i$ is the gauge field of $\SU(\ell_i)$.

In the coordinate system in \eqref{construction},
the generator of $\SO(2)_T$ (of the directions 45)
and that of $\SO(2)_R$ (of the directions 67) are 
$\partial_\beta-\partial_\chi$ and  $-\partial_\chi$, respectively.
Therefore, the gauge fields $A_{T,R}$ on the $AdS_5$ corresponding to the rotations $\SO(2)_{T,R}$ enter into the configuration by replacing \begin{equation}
d\beta \to d\beta + A_T, \qquad
d\chi \to d\chi-  A_T - A_R.  \label{gauging}
\end{equation}
Plugging this into \eqref{CS} and  integrating over $S^2$, 
we obtain the Chern-Simons terms among $\SO(2)_{T,R}$ and $G_Y$ we wanted to consider. 

We conclude that the 't Hooft anomalies of 
the flavor symmetries $G_Y$ of the defect of type $Y$
are encoded in the anomaly polynomial
\begin{equation}
\sum_i \left[ (-\eta_i+\lambda_{\eta_i}) \frac{F_T}{2\pi} + \lambda_{\eta_i} \frac{F_R}{2\pi} \right]  \frac{1}{8\pi^2}\tr F_i\wedge F_i.\label{6danomaly}
\end{equation}
Here, $F_T$, $F_R$ and  $F_i$ are the external gauge fields coupled to $\SO(2)_T$, $\SO(2)_R$ and $\SU(\ell_i) \subset G_Y$, respectively.  $\U(1)$ parts are subtler.

Here we need a standard disclaimer saying that the gravity solutions can only be trusted when the curvature is small except at the $\bC^2/\bZ_n$ singularities; but the anomaly polynomial is a robust quantity and we believe \eqref{6danomaly} is correct as is, for all $Y$.

\section{Conclusions}\label{conclusions}

%\section{Synthesis}\label{synthesis}

The 4d theory discussed in Sec.~\ref{4} is obtained by compactifying the directions 45 in the brane system \eqref{construction} on a Riemann surface $C$. Then the $\U(1)_R$  symmetry of the resulting $\cN=2$ SCFT is just the rotation $\SO(2)_R$.
This way the anomaly polynomial \eqref{6danomaly} explains the anomaly \eqref{4danomaly}.
Essentially the same analysis was already given in \cite{Gaiotto:2009gz}.

The 2d theory discussed in Sec.~\ref{2} is conjectured to arise by compactifying the directions 2345 in the brane system \eqref{construction} on $\bR^4_{\epsilon_1,\epsilon_2}$. Here $\epsilon_1$ is the equivariant parameter for the rotation of directions 23,
and $\epsilon_2$ is that for directions 45. 
To preserve the supersymmetry, we perform the standard Donaldson-Witten twist, setting
the equivariant parameter for the directions 67 to be $-\epsilon_1-\epsilon_2$.

We  perform the integration of the anomaly polynomial as in \cite{Alday:2009qq}: the essential point is that the Chern root $F/2\pi$ of the $\bR^2$ bundle is replaced by the corresponding equivariant parameter, and $\int_{\bR^2_{\epsilon}} 1=1/\epsilon$.
Thus, the integral of $F_T$, $F_R$ in \eqref{6danomaly} along the directions 23 gives \begin{equation}
\int_{\bR^2_{\epsilon_1}} F_T = \frac{\epsilon_2}{\epsilon_1},\qquad
\int_{\bR^2_{\epsilon_1}} F_R = -\frac{\epsilon_1+\epsilon_2}{\epsilon_1}.
\end{equation}
Then the anomaly  \eqref{6danomaly} reduces to \begin{equation}
\sum_i (-\lambda_{\eta_i} - \eta_i\frac{\epsilon_2}{\epsilon_1} ) \frac{1}{8\pi^2}\tr F_i\wedge F_i.
\end{equation} 
This reproduces the level of the current algebras \eqref{2dlevel} by taking \begin{equation}
b^2=\frac{\epsilon_2}{\epsilon_1}.
\end{equation}
This identification between $b^2$ and the ratio of the equivariant parameters is the standard one \cite{Alday:2009aq}.

The fact that these levels are reproduced from the anomaly of the defects  provides a check of the conjecture that this compactification does give rise to a 2d theory with the W-symmetry associated to the quantum Drinfeld-Sokolov reduction for a general embedding $\rho_Y:\SU(2)\to\SU(N)$.

As for future directions of research, we first need to complete the determination of the anomaly polynomial of the codimension-two defect, preferably without using holography.
In this paper, we only studied $\SO(2)_{R,T}G_Y^2$ involving the non-Abelian part of $G_Y$.
 There are also anomalies cubic in $\SO(2)_{R,T}$ and  ones involving gravitational contributions. 
Upon compactification, they would determine $a$ and $c$ of the 4d $\cN=2$ theory,
and also the central charge of the W-symmetry of the 2d theory.
It would also be interesting to generalize the analysis of the anomaly  to the defects of 6d $\cN=(2,0)$ theory of type $D$ and $E$.  These are left as exercises to the reader.

A more distant project would be to understand  the Drinfeld-Sokolov reduction itself from the point of view of the codimension-two defect. As reviewed in Sec.~\ref{2}, general W-algebras arise from the reduction of the affine $\hat\SU(N)$ symmetry. Similarly, any defect labeled by a Young diagram can be obtained by going to the Higgs branch of the defect with $\SU(N)$ symmetry \cite{Gaiotto:2009we,Benini:2009gi}.  There is a possibility that the latter leads to the former upon compactification.

\section*{Acknowledgments}
The author thanks Ruben Minasian for fruitful discussions which led to the ideas expressed in the paper.
He thanks Davide Gaiotto for pointing out incorrect statements in Sec.~\ref{5} in version 1 of this note.
He also thanks Jaume Gomis, Simeon Hellerman, Takuya Okuda, Shigeki Sugimoto for discussions.  
He is supported in part by World Premier International Research Center Initiative (WPI Initiative),  MEXT, Japan through the Institute for the Physics and Mathematics of the Universe, the University of Tokyo.
He is also supported in part by NSF grant PHY-0969448  and by the Marvin L. Goldberger membership through the Institute for Advanced Study.

\appendix 

\section{More on the brane constructions}\label{appendix}
Here we describe in more detail the configurations of the branes, which give rise to the codimension-two defects in the 6d $\cN=(2,0)$ theory in the low energy limit.  We do not discuss all of the possible supersymmetric defects.\footnote{For other type of defects, see e.g.~\cite{Drukker:2009tz,Drukker:2009id,Chen:2010jga,Drukker:2010jp,Hosomichi:2010vh}.}

Let us start by considering the following two IIA configurations:
%\begin{subequations}
\begin{align}
&
\begin{array}{l||cc|cc|cc|c|ccccc}
&0&1&2&3&4&5&6&7&8&9 \\
\hline
\hbox{NS5} & - & - & -& -& - & -&& \\
\hbox{D4} & - & - & -& -&  &  & - &&\\
\hbox{D4'} & - & - & &  &  &  & -  & - & -
\end{array}
\\[1em]
&
\begin{array}{l||cc|cc|cc|c|ccccc}
&0&1&2&3&4&5&6&7&8&9 \\
\hline
\hbox{NS5} & - & - & -& -& - & -&& \\
\hbox{D4} & - & - & -& -&  &  & - &&\\
\hbox{D4'} & - & - & &  & -& -& - & & 
\end{array}
\end{align}
%\end{subequations}

In both cases, we put a few NS5s, and suspend $N$ D4s between them,
producing $\cN=2$ gauge theory on the directions 0123.
The D4'-brane then creates the surface operator on the directions 01. 

They lift to the following M-theory configurations, respectively:
%\begin{subequations}
\begin{align}
&
\begin{array}{l||cc|cc|cc|cc|ccccc}
&0&1&2&3&4&5&6&11&7&8&9 \\
\hline
\hbox{M5}^\circ & - & - & -& -& - & - &&&&  \\
\hbox{M5} & - & - & -& -&  &  & - &- &&\\
\hbox{M5'} & - & - & &  &  &  & -  & - & - &-&
\end{array} \label{A}
\\[1em]
&
\begin{array}{l||cc|cc|cc|cc|ccccc}
&0&1&2&3&4&5&6&11&7&8&9 \\
\hline
\hbox{M5}^\circ & - & - & -& -& - & - &&&&  \\
\hbox{M5} & - & - & -& -&  &  & - &- &&\\
\hbox{M5'} & - & - & &  & -& -& -  & - & &&
\end{array} \label{B}
\end{align}
%\end{subequations}

The bulk 6d theory is on the worldvolume of $N$ M5-branes extending along the directions 01236 and 11.
The directions 6-11 are compactified on a Riemann surface $C$.
The defects created by M5$^\circ$ change the 4d theory on directions 0123,
and the defects created by M5$'$, which are surface operators of that 4d theory,
change the theory on $C$. 

Let us study which supercharges are preserved. 
Denote the spinor components by 
$(\pm\pm\pm\pm\pm)$ where $\pm$'s stand for the eigenvalue of $\Gamma^{01}$, $\Gamma^{23}$, $\Gamma^{45}$, $\Gamma^{6,11}$, $\Gamma^{78}$ respectively. An M5 along $012345$ preserves spinors with $\Gamma^{01}\Gamma^{23}\Gamma^{45}=+1$, etc.
Then it follows that the configuration \eqref{A} preserves
%\begin{subequations}
\begin{equation}
\begin{array}{ll}
(+++++), &
(+----),\\
(-+--+),&
(--++-),
\end{array} \label{Q} 
\end{equation} giving rise to an $\cN=(2,2)$ surface operator on the direction 01. 
Meanwhile, the configuration \eqref{B} preserves
\begin{equation}
\begin{array}{ll}
(+++++),&
(+---+),\\
(++++-),&
(+----),
\end{array}
\end{equation}
%\end{subequations}
giving rise to an $\cN=(4,0)$ surface operator.  
In papers \cite{Alday:2010vg,Kozcaz:2010yp} dealing with the surface operator coming from the codimension-2 defects, it was not clearly stated which of the setup was taken. 
Instead it was implicitly assumed that the physics of the BPS sector reduces to the supersymmetric quantum mechanics on the moduli space of instantons with prescribed singularities. 
The moduli space was K\"ahler and not hyperk\"ahler. 
This means that in those papers the $\cN=(2,2)$ operator, or equivalently the setup \eqref{A} was implicitly chosen.

To put the 4d theory preserving supersymmetry on directions 0123 on a nontrivial manifold (including $\bR^4_{\epsilon_1,\epsilon_2}$),
we need to perform the standard Donaldson-Witten twist.
This sets $A_{78}=-A_{01}-A_{23}$, where $A_{ij}$ is the affine or R-symmetry connection corresponding to the rotation of the directions $i$ to $j$.

In the main part of the note, we discussed the manifestation of the defect in 4d theory and in 2d theory.
For the former, the configuration in \eqref{construction} is obtained by dropping M5' in \eqref{A}, and redefining the coordinates as in 
\begin{equation}
\begin{array}{c||cccc|cc|cc|cccc}
\eqref{construction} & 0 &1 &  2&  3&  4&  5&  6&  7&  8&  9& 10 \\
\hline
\eqref{A} & 0 & 1& 2 & 3 & 6 & 11 & 4 & 5 & 7 & 8 &9
\end{array}.
\end{equation}
For the latter, the configuration \eqref{construction} is obtained by dropping M5$^\circ$ in \eqref{A} and  rearranging the directions via \begin{equation}
\begin{array}{c||cc|cccc|cc|cc|cc}
\eqref{construction} & 0 &1 &  2&  3&  4&  5&  6&  7&  8&  9&  10 \\
\hline
\eqref{A} & 6 & 11 & 0& 1 & 2 & 3 & 7 & 8 & 4 & 5 & 9
\end{array}.
\end{equation}
After this latter operation, M5$^\circ$ (or more general defects labeled by any Young diagram $Y$) creates a state in the 2d CFT living in the direction 01. The way the type of $Y$ is manifested in the state was studied in \cite{Kanno:2009ga,Kanno:2010kj,Drukker:2010vg}.

One final comment is that the surface operator given in \eqref{A} preserves the same supercharges as the surface operator coming from an M2, treated e.g.~in \cite{Alday:2009fs,Tan:2009qq,Kozcaz:2010af,Dimofte:2010tz,Taki:2010bj,Awata:2010bz}.
This has the setup 
\begin{equation}
\begin{array}{l||cc|cc|cc|cc|ccccc}
&0&1&2&3&4&5&6&11&7&8&9 \\
\hline
\hbox{M5} & - & - & -& -& - & - &&&&  \\
\hbox{M5} & - & - & -& -&  &  & - &- &&\\
\hbox{M2} & - & - & &  &  &  &   & &  &&-
\end{array}.
\end{equation}
The supersymmetry preserved by M2 has eigenvalue $+1$ under the product of $\Gamma$'s transverse to M2.
Then the  components of supercharges preserved by the total system are exactly the ones in \eqref{Q}.

\bibliographystyle{utphys}
\baselineskip=.89\baselineskip
\bibliography{w}{}
%%list W-algebra

\end{document}